\documentstyle[12pt]{article}
\begin{document}
\baselineskip=24pt
\begin{center}
{\large Effective action for the Kondo lattice model.\\
New approach for $S=1/2$.}
\\
F.~Bouis$^{a}$ and M.N.~Kiselev$^{a,b}$\\
{\it $^a$Laboratoire L\'eon Brillouin - CE-Saclay 91191,
Gif-sur-Yvette, France\\
 $^b$Russian Research Center "Kurchatov Institute",
123 182 Moscow, Russia}
\end{center}
\begin{abstract}
In the partition function of the Kondo lattice (KL), spin matrices are exactly replaced by bilinear combinations of Fermi operators with the purely imaginary  chemical potential $\lambda=-i\pi T/2$ (Popov representation). 
This new representation of spin operators allows one to introduce 
new Green Functions (GF) with Matsubara frequencies $\omega_n=2\pi T(n+1/4)$ for $S=1/2$. A simple temperature 
diagram technique  is constructed with the path integral method.
This technique is standard and does not contain
the complicated combinatoric rules characteristic of most of the known
variants of the diagram techniques for spin systems. 
The effective action for the almost antiferromagnetic KL problem is derived.
\end{abstract}

\newpage 
Many systems in statistical physics are described by Hamiltonians containing spin matrices. Unfortunately, the diagrammatic perturbation theory for spin systems is complicated. Many variants are based on different representations of the spin matrices by Bose or Fermi operators. However, unphysical states always arise leading to constraints and complication of the Feynman codex.
In this paper we construct a simple diagrammatic technique (DT) for spin-$\frac{1}{2}$ that differs from the known techniques in the form of the GF, but which is standard in other respects, does not contain the complicated combinatoric rules characteristic of spin systems and permits one to take into account the constraints rigorously.

It is indeed possible to replace exactly spin-$\frac{1}{2}$ matrices by bilinear combinations of Fermi operators: 
\begin{equation} \label{representation}
\sigma_{f{\bf i}}^z  \rightarrow  S_{f{\bf i}}^z= \frac{1}{2}(a_{{\bf i}\uparrow}^+ a_{{\bf i}\uparrow} - a_{{\bf i}\downarrow}^+ a_{{\bf i}\downarrow}) \, ; \, 
\sigma_{f{\bf i}}^+  \rightarrow  S_{f{\bf i}}^+= a^+_{{\bf i}\uparrow} a_{{\bf i}\downarrow} \, ; \,
\sigma_{f{\bf i}}^-  \rightarrow   S_{f{\bf i}}^- =a^+_{{\bf i}\downarrow} a_{{\bf i}\uparrow},
\end{equation}
by the basic formula shown in Ref. \cite{popov} (see also Ref. \cite{oppermann}): 
\begin{equation}
Z=\mbox{Sp } e^{-\beta H} =i^N \mbox{Sp } e^{-\beta(H_f+\frac{i\pi}{2\beta}N)},
\end{equation}
where $H_f$ is obtained from H by the replacement (\ref{representation}), and $N=\displaystyle\sum_{{\bf i}\sigma} a^+_{{\bf i}\sigma}a_{{\bf i}\sigma}$. There is no constraint but the purely imaginary chemical potential of pseudofermions $\lambda=-i\pi T/2$ leads to the mutual cancellation of the unphysical states. 

We analyze here the KL model which is a periodic lattice of 
magnetic atoms modeled by $f$-orbitals in a metallic background:  \begin{equation}
{\cal H}_{KL}=-\sum_{{\bf ij},\sigma} (t_{\bf ij}+\mu) \Psi^+_{{\bf i}\sigma} \Psi_{{\bf j}\sigma}
+J_{sf}\sum_{\bf i} \Psi^+_{{\bf i}\sigma} \sigma\Psi_{{\bf i}\sigma'} {\bf S}_{f\,{\bf i}}+g\sum_{\bf i}(H+he^{i{\bf R_i}{\bf Q}})S^z_{f {\bf i}}.
\label{h1}
\end{equation}
We add a uniform ($H$) and a staggered ($h$) magnetic field ($g=\mu_B{\em g}_L$, where $\mu_B$ is the Bohr magneton and ${\em g}_L$ is the Land\'e factor). We consider a simple cubic lattice with the notation ${\bf Q}={\bf Q}_{AF}=(\pi,\pi,\pi)$.
Using Popov representation of spins, the ratio of the partition function of the interacting system 
 to the partition function of the corresponding free system can be represented in the form of functional integrals as follows:
\begin{equation}
Z/Z_0=\frac{\int D\nu \exp\left[S-\frac{i\pi}{2\beta}\int_0^\beta
d\tau\sum_{\bf i}\bar{a}_{{\bf i},\alpha}(\tau)a_{{\bf i},\alpha}(\tau)\right]}{
\int D\nu \exp\left[S_0-\frac{i\pi}{2\beta}\int_0^\beta
d\tau\sum_{\bf i}\bar{a}_{{\bf i},\alpha}(\tau)a_{{\bf i},\alpha}(\tau)\right]},
\label{pf}
\end{equation}
where the Euclidean action for the KL model is 
\begin{equation}
S=\int_0^\beta d\tau \{\sum_{\bf i}[\bar{\Psi}_{{\bf i},\sigma}\partial_\tau\Psi_{{\bf i},\sigma}
+\bar{a}_{{\bf i},\alpha}\partial_\tau a_{{\bf i},\alpha}] - {\cal H}_{KL}(\tau)\}.
\label{s1}
\end{equation}
We note by $D\nu$ the integration over the
  anticommuting Grassmann variables $\Psi_\sigma, a_\alpha$. By making the replacement
$a_{{\bf i},\alpha}(\tau) \to a_{\bf i}^{\alpha}(\tau)\exp(\frac{i\pi}{2\beta}\tau)$ etc, which cancels the last term in both exponents in numerator and denominator of Eq. (\ref{pf}), we come to the following boundary conditions for Grassmann fields:
$
\Psi_\sigma(\beta)=-\Psi_\sigma(0),\;\;
\bar{\Psi}_\sigma(\beta)=-\bar{\Psi}_\sigma(0),\;\;
\bar{a}^\alpha(\beta)=-i\bar{a}^\alpha(0),\;\;
a^\alpha(\beta)=ia^\alpha(0).
$
Going over to the momentum representation for all Grassmann variables
and assuming ${\bf s}_{sk}=\displaystyle\sum_p\bar{\Psi}_{p+k}\sigma\Psi_{p}$ we obtain:
\begin{equation}
S=\sum_{ k}\bar{\Psi}_{k,\sigma}G_0^{-1}\Psi_{k,\sigma} +
\sum_{ p}\bar{a}_{p}^{\alpha}{\cal G}_0^{-1}a_{p}^{\alpha}
+ J_{sf} \sum_{k} {\bf s}_{s \, k} {\bf S}_{f \, -k} +
\frac{1}{2}hg
\sum_{ k}\bar{a}_{k}^{\alpha}\sigma^za_{k+Q}^{\alpha'},
\label{s2}
\end{equation}
where the inverse GF of $\Psi$-fields is $G_0^{-1}=i2\pi T(n+1/2)-\varepsilon_{\bf k} +\mu$ with dispersion 
$\varepsilon_{\bf k} = \displaystyle
-\sum_{\bf \delta} t_{\bf i,i+\delta} e^{i{\bf \delta k}}$
and the inverse GF of $a^\alpha$ Grassmann fields is ${\cal G}_{0\alpha}^{-1}=i2\pi T(m+1/4)-1/2gH\sigma^z_{\alpha\alpha}$ with unusual Matsubara frequencies. Note that Popov representation can be used for spins $S=1$ as well. In this case the frequencies are shifted to $\omega_m=2\pi T (m+1/3)$. Moreover the method has been also extended to arbitrary spin in Ref. \cite{oppermann}.

We now confine ourselves to the limiting case $T_N \sim T_K \sim T_0$ 
\cite{Don} assuming the same energy scale for antiferromagnetic (AF) and Kondo correlations. It allows us to integrate over the fast $\Psi$ fields with energies $\varepsilon \sim \mu \gg T_0$ using the bare electrons GF. We can also integrate over the fast 
 fields $a^\alpha$ ($\omega \gg T_0$) taking into account
one-site Kondo renormalization of vertices ($J_{sf}\to {\cal J}_{sf}$) and self-energy parts ($G_0 \to G$) \cite{KKM}. As a result, a simple DT is constructed. Contrary to other DT (see, e.g. \cite{divers}-\cite{read}) the constraint on the spin subsystem is taken into account exactly. The new action which is written in terms of slow $\Psi$ and $a^\alpha$ variables contains an additional AF Heisenberg interaction between spins due to the indirect RKKY exchange \cite{KKM}-\cite{divers}:
\begin{equation} 
\label{full}
S_{eff}=\sum_{k}\bar{\Psi}^{slow}_{k,\sigma}G^{-1}\Psi^{slow}_{k,\sigma}
+ \sum_{k} {\cal J}_{sf} {\bf s}^{slow}_{s \, k} {\bf S}_{f \, -k} +S_H
\label{s4}
\end{equation}
The last term in (\ref{s4}) can be analyzed separately and represented by auxiliary three-component Bose fields $\phi^\gamma(k)$ \cite{popov}:  
\begin{equation} \label{former}
S_H = \sum_{k_1k_2 \sigma} \bar{a}_{k_1}^{\alpha}[{\cal G}_0^{-1}\delta_{k_1,k_2}+\sigma^z h\delta_{k_1+Q,k_2}]a_{k_2}^{\alpha} 
 -\sum_{k}[\frac{1}{2}\phi^3_k S^z_{-k} + \eta^*_{k} S^-_{k}+ \eta_{k} S^+_{k}]  +S_0[{\bf \phi}]
\end{equation}
with the following notation: $S_0  =  -1/4 \displaystyle\sum_{k} (\Gamma^{RKKY}_{ k})^{-1} \phi^\gamma_{k} \phi^\gamma_{-k}$ ,
$\eta^*_{k}  =  (\eta_k)^*$
 and $\eta_{k}  =  (\phi^1_{k}-i\phi^2_{k})/2$. 
In the case $T_K \ll T_N$ only magnetic terms in the effective action are important. We note by W the matrix of the quadratic form in $a^\alpha$ variables. Integrating over all $a^\alpha$ fields one can find the nonpolynomial action of the AF Heisenberg model in terms of Bose fields \cite{BouKi}:
\begin{equation}
S_H  =  S_0[\phi^3,\eta] +\log \det W[\phi^3,\eta].
\label{non}
\end{equation}
In the case considered, namely $T_N \sim T_K \sim T_0$ the procedure of derivation of the effective action is a little bit complicated. Taking into account the second term in (\ref{full}) one has to replace $\phi^\gamma \rightarrow \phi^\gamma - 2{\cal J}_{sf} s_{s,k}^{\gamma \, slow}$ in (\ref{non}). As a result, the effective action can be rewritten as follows:
\begin{equation} 
\label{result}
S_{eff}  =  \sum_{k}\bar{\Psi}^{slow}_{k,\sigma}G^{-1}\Psi^{slow}_{k,\sigma}+
S_0[\phi^3,\eta] +\log \det W[\phi^\gamma_k-2{\cal J}_{sf} {\bf s}^{\gamma \, slow}_{s \, k}].
\end{equation}
 Equations (\ref{full}-\ref{result}) are the key result of the present work.
This effective action describes the low energy properties of the KL model. The last term in (\ref{non}) takes into account the mutual influence of conduction electrons and spins. Magnetic instabilities of both kind of electrons could then  be easely analysed.

Let us concentrate on the former problem (\ref{former}).
The spin subsystem undergoes a phase transition with $T_c=T_0$ corresponding to the appearance of a nonzero staggered magnetization $\rho$ as $h\to 0$. This problem is related to the Bose-condensation of the field
$
\phi^3_{k}=\tilde{\phi}^3_{k}+\rho (\beta N)^{\frac{1}{2}} \delta_{\bf k,Q}\delta_{\omega}
$
and in one-loop approximation results in the usual mean-field equation  for AF order parameter \cite{BouKi} in the presence of Kondo-scattering processes
\cite{KKM}. Note that a magnetic transition in the localized system may induce a
magnetic transition in the itinerant system.

Taking into account the compensation equation \cite{KKM},\cite{BouKi} and calculating the
$\log \det W[\phi^3,\eta]$ approximately by the method of stationary phase the following expression for the spin subsystem effective action can be obtained:
\begin{equation}
S^{eff}_H=\sum_{k}\eta^*_k \chi_t^{-1}\eta_k+
\sum_{k}\phi^{3*}_k \chi_l^{-1}\phi^3_{-k}-
1/4\sum_{k} (\Gamma^{RKKY}_{\bf k})^{-1} \phi^\gamma_{k} \phi^\gamma_{-k},
\label{s3}
\end{equation}
where $\chi_t$ and $\chi_l$ are transverse and longitudinal susceptibilities respectively. As usual, the transverse susceptibility describes the AF magnons excitations. At the temperature range $T>T_0$ when the condensate solution is
absent the effective action has the same form except that the transverse and longitudinal susceptibilities describe the paramagnon excitations which can result in some untrivial effects in heavy-fermion compounds \cite{pyka}. These excitations 
introduce a new energy scale corresponding to the critical behaviour.

Summarizing, we constructed a simple diagrammatic technique which allows one to analyze the effective action of the KL model when the energy scales for AF and Kondo correlations are the same. This effective action describes the slow electron subsystem interacting with the spin fluctuations of either magnon or paramagnon type.

The support of the grant RFBR 98-02-16730 (MNK) is acknowledged.

\end{document}